\documentclass[%
 reprint,
 amsmath,amssymb,
 aps,
]{revtex4-2}

\usepackage{graphicx}
\usepackage{dcolumn}
\usepackage{bm}
\usepackage{subfigure}
\usepackage{hyperref}
\usepackage{tikz}

\begin{document}

\preprint{APS/123-QED}

\title{Suppression of wall modes in rapidly rotating Rayleigh-B\'enard convection\\ by narrow horizontal fins}

\author{Louise Terrien}
 \email{louise.terrien@ens-paris-saclay.fr}
\author{Benjamin Favier}%
 \email{benjamin.favier@cnrs.fr}
\affiliation{%
Aix Marseille Univ, CNRS, Centrale Marseille, IRPHE, Marseille, France}
\author{Edgar Knobloch}%
 \email{knobloch@berkeley.edu}
\affiliation{Department of Physics, University of California at Berkeley, Berkeley, California 94720, USA}

\date{\today}

\begin{abstract}
The heat transport by rapidly-rotating Rayleigh-B\'enard convection is of fundamental importance to many geophysical flows. Laboratory measurements are impeded by robust wall modes which develop along vertical walls, significantly perturbing the heat flux. We show that narrow horizontal fins along the vertical walls efficiently suppress wall modes ensuring that their contribution to the global heat flux is negligible compared with bulk convection in the geostrophic regime, thereby paving the way for new experimental studies of geophysically relevant regimes of rotating convection.
\end{abstract}

\maketitle



Geostrophic turbulence is of fundamental importance to rapidly rotating flows satisfying geostrophic balance.
This state can be realized in laboratory experiments usually performed in tall (to reach high Rayleigh numbers) and thin (to reduce the Froude number measuring centrifugal effects) cylinders \cite{Ecke2014,cheng2015,aurnou_bertin_grannan_horn_vogt_2018,cheng2018heuristic,cheng2020}.
These experiments are plagued by the presence of robust wall modes localized at vertical boundaries \cite{zhongES,goldstein,horn2017prograde,favier2020robust}.
In thin cylinders these modes contaminate bulk heat flux measurements \citep{kunnen,bodenschatz,kunnen21,zhang2021boundary,ecke2022connecting,ecke2022turbulent} degrading the ability to study geostrophic turbulence in the laboratory.

The shape and roughness of the solid boundary are often used in fluid mechanics \citep{jimenez2004} to control or delay (resp. favor) undesirable (resp. desirable) bifurcations, e.g. in Taylor-Couette flows \citep{cadot1997,berg2003,zhu2016direct} or Rayleigh-B\'enard convection \citep{shen1996,ciliberto1999,wagner2015heat,xie_xia_2017}.
Unfortunately, wall modes cannot be eliminated by inserting vertical barriers into the flow \citep{favier2020robust}, a property reminiscent of topologically-protected edge states \citep{kane2014topological,irvine,yang2015,souslov2019}.
In the present work we show that this is not the case for {\it horizontal} barriers.
Our detailed results indicate that the insertion of narrow horizontal fins along the lateral boundary provides efficient wall mode suppression, thereby enabling laboratory studies of a key geophysical process.  


We consider an incompressible fluid with constant kinematic viscosity $\nu$ and thermal diffusivity $\kappa$ inside a rectangular box of dimension $L_x\times L_y\times H$, heated from below and cooled from above.
The temperatures at the bottom and top are fixed and the vertical walls perpendicular to the $x$-axis are thermally insulating; all walls are impenetrable and no-slip.
The domain is periodic in the $y$ (or azimuthal) direction and the system rotates around the $z$-axis at a constant rate $\bm{\Omega}=\Omega \bm{e_z}$. Gravity is downward, $\bm{g}=-g \bm{e_z}$.
Laboratory experiments employing liquids are well described by the Boussinesq approximation with constant density except in the buoyancy term.
Using $1/(2\Omega)$ as the unit of time and the depth $H$ as the unit of length, the dimensionless equations are:
\begin{align}
\label{eq:momentum}
& \frac{\partial \bm{u}}{\partial t}+\bm{u}\cdot\nabla \bm{u} =\! -\nabla p -\bm{e_z}\!\times\!\bm{u}+\frac{RaE^2}{Pr}T\bm{e_z}\!+\!E\nabla^2\bm{u} \\
\label{eq:incomp}
& \nabla\cdot\bm{u}=0 \\
\label{eq:temp}
& \frac{\partial T}{\partial t}+\bm{u}\cdot\nabla T = \frac{E}{Pr}\nabla^2 T,
\end{align}
where $\bm{u}$ is the velocity, $T$ the temperature and $p$ the pressure. Centrifugal effects are neglected \cite{horn2018regimes}.
The system is characterized by three dimensionless parameters: the Rayleigh number $Ra=\alpha g \Delta T H^3/(\nu\kappa)$, the Ekman number $E=\nu/(2\Omega H^2)$ and the Prandtl number $Pr=\nu/\kappa$.
Here $\alpha$ is the thermal expansion coefficient and $\Delta T$ is the imposed temperature difference across the layer.
In the following we set $Pr=1$.

\begin{figure}
    \centering
    (a)\includegraphics[width=0.169\textwidth]{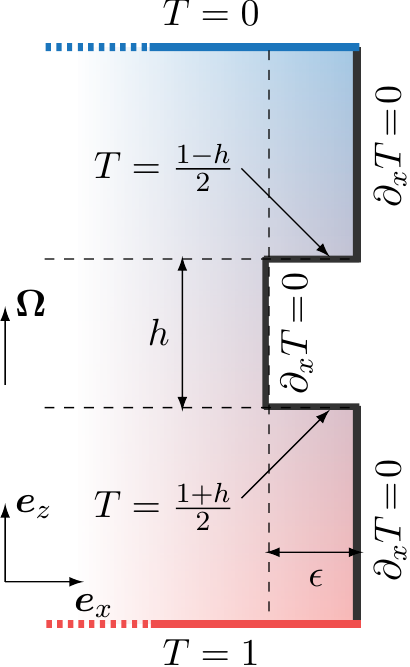}\hspace{8mm} (b)\includegraphics[width=0.15\textwidth]{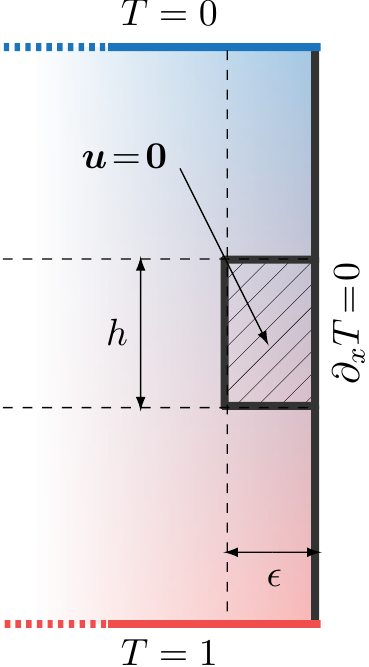}
    \caption{Transverse section of the domain in the $(x,z)$-plane, focusing on one side of the domain. All boundaries are no-slip with periodic boundary conditions in the $y$ direction. (a) Imposed temperature barrier: all vertical boundaries are thermally insulating with imposed temperatures on the two horizontal boundaries equal to those in the equilibrium background. (b) Conducting barrier: the exterior vertical boundary is insulated but heat can diffuse through the barrier with the same diffusivity as in the liquid.}
    \label{barrier}
\end{figure}

We enrich this otherwise classical problem by introducing identical horizontal barriers or fins along both vertical walls.
These barriers are invariant in the $y$ direction and of rectangular cross-section with horizontal width $\epsilon$ and vertical extent $h$ centered on the midheight $z=1/2$. See Fig.~\ref{barrier}.

We consider two types of boundary conditions on this obstacle.
In the first case, the vertical side is insulating while the horizontal sides are maintained at fixed temperatures equal to those in the local equilibrium profile $T(z)\!=\!1\!-\!z$.
For a barrier of height $h$ centered on $z\!=\!1/2$, the imposed temperature is $T\!=\!(1\!-\!h)/2$ at the upper surface and $T\!=\!(1\!+\!h)/2$ at the lower surface [Fig.~\ref{barrier}(a)].
While unrealistic from an experimental point of view, this is a well-posed problem that prevents the development of baroclinic flows around the intrusion (which would develop for a fully insulating barrier) and provides a consistent stable equilibrium around which perturbations can be studied.
The second type of barrier conducts heat and we impose an insulating boundary condition on the original sidewall [Fig.~\ref{barrier}(b)].
To prevent the emergence of baroclinic flows, we assume that the thermal diffusivity inside of the barrier is the same as that of the fluid flowing around it.
The presence of baroclinic flows complicates the analysis in either case but does not lead to a qualitative change in our conclusions \citep{supp}.
The simple question we ask here is the following: for which values of $h$ and $\epsilon$ are wall modes suppressed for given $E$ and $Ra$?

We solve Eqs.~\eqref{eq:momentum}-\eqref{eq:temp} using the spectral-element code Nek5000\footnote{NEK5000 Version 19.0. Argonne National Laboratory, Illinois. Available: \url{https://nek5000.mcs.anl.gov}.} \citep{fischer1997overlapping}. The mesh is composed of up to 17280 hexahedral elements and we use a polynomial order up to $N=13$ including dealiasing. The mesh is refined close to the horizontal (resp. vertical) boundaries of the domain in order to properly resolve Ekman layers (resp. wall modes). Numerical convergence of the results has been checked by gradually increasing the polynomial order for a given number of elements. The equations are solved as an initial value problem, even though some of our results concern the exponential growth rate of perturbations. This approach is appropriate for complex domains such as ours given that the same procedure allows us to study the nonlinear and indeed turbulent state of the system.


We first fix the Ekman number at $E=10^{-4}$, a small enough value to clearly isolate wall modes from regular bulk convection.
The Rayleigh number is fixed to twice the critical value for the onset of wall modes, $Ra\!=\!6.8\!\times\!10^{5}\!\approx\!2Ra^{\textrm{wall}}_c$ where $Ra_c^{\textrm{wall}}$ is the critical $Ra$ at which wall modes first set in \citep{busse1993,zhang_liao_2009}.
The period in the $y$ direction is chosen to be twice the most unstable wavelength predicted by linear theory for the case without a barrier, a trade-off between accuracy and numerical cost.
We have checked that increasing $L_x$ did not significantly affect the estimated growth rate.
The distance between the two lateral walls is chosen to be approximately 20 times the typical width of the wall modes \cite{busse1993}, ensuring that the two wall modes on either side do not interact.
Our simulation parameters are summarized in \citep{supp}.

\begin{figure}
    \centering
    (a)\includegraphics[width=0.41\textwidth]{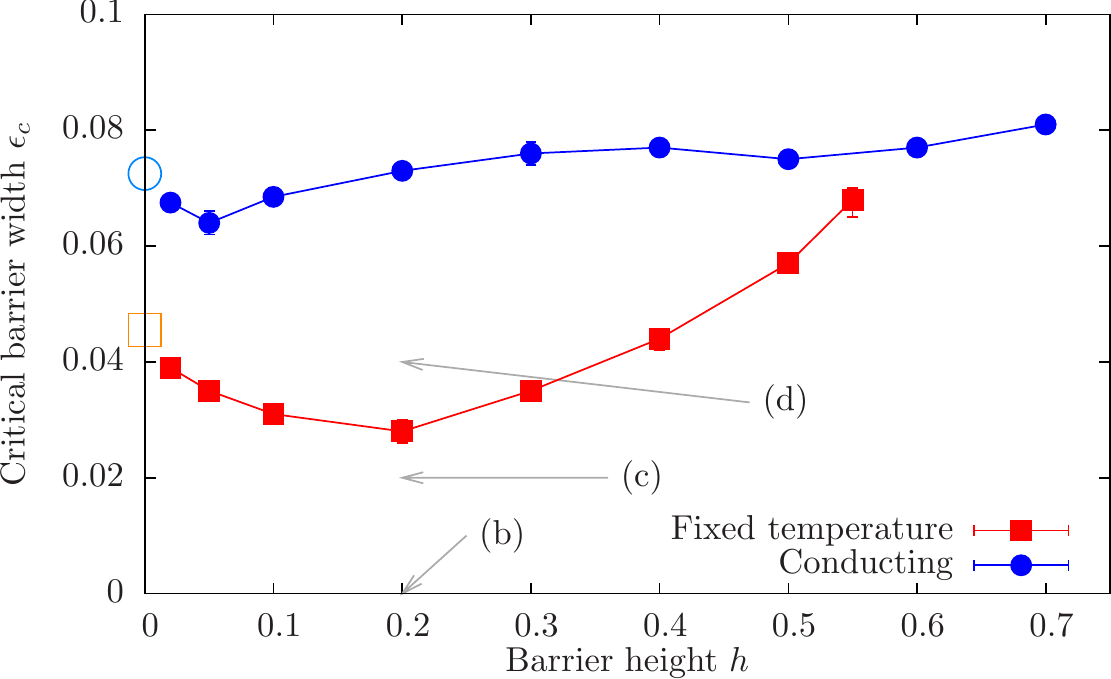}\\ \vspace{1mm}
    (b)\includegraphics[width=0.13\textwidth]{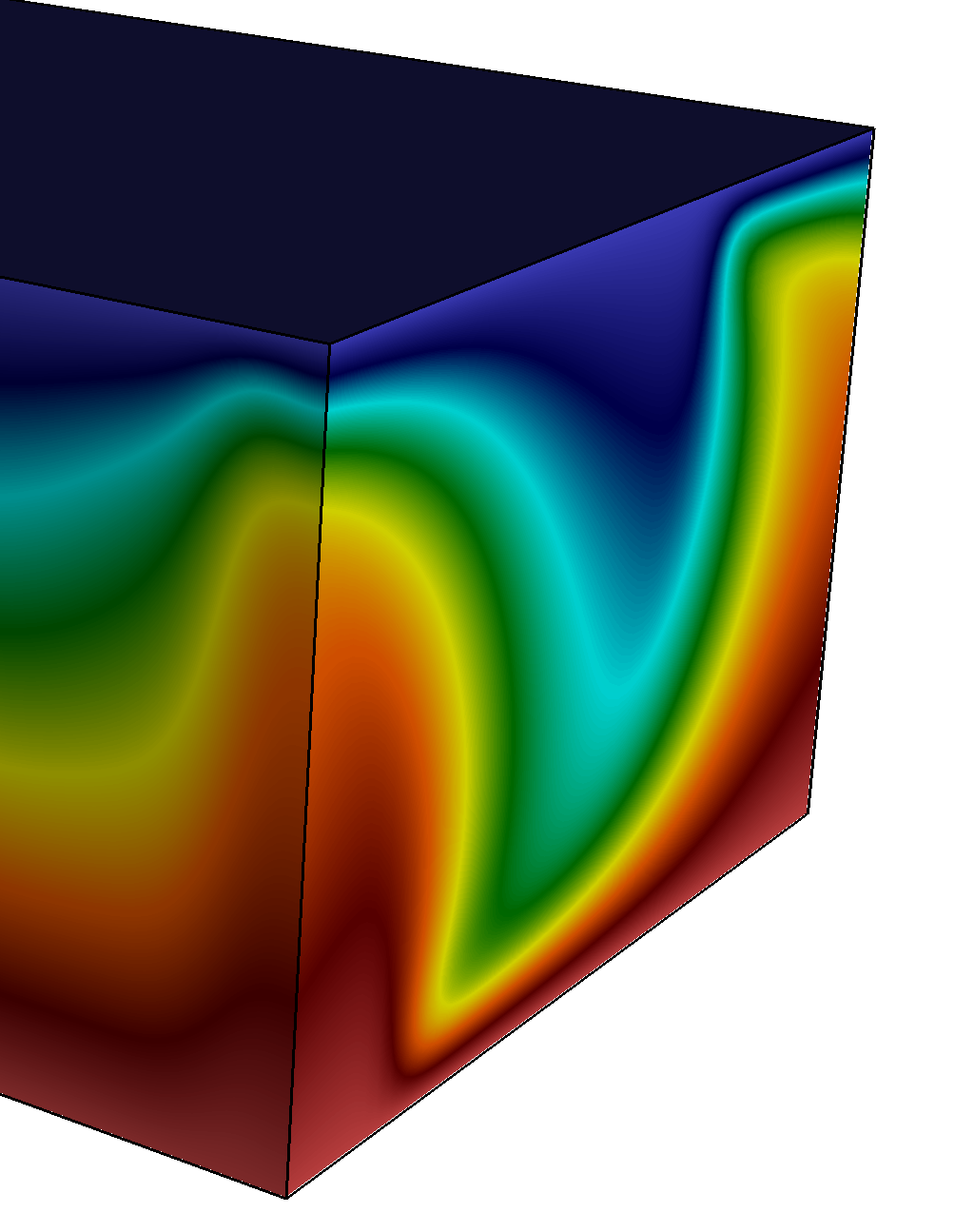}\hfill
    (c)\includegraphics[width=0.13\textwidth]{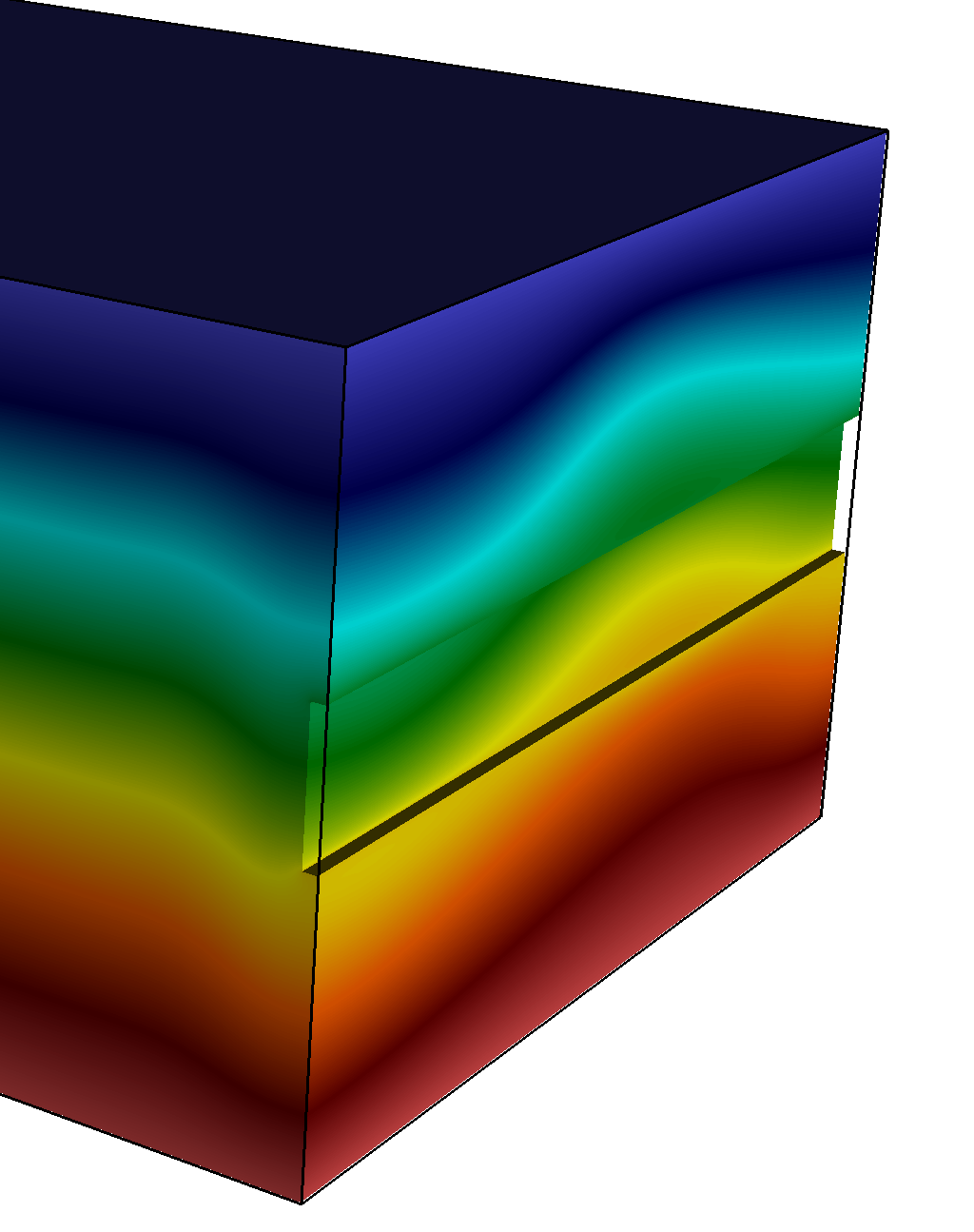}\hfill (d)\includegraphics[width=0.13\textwidth]{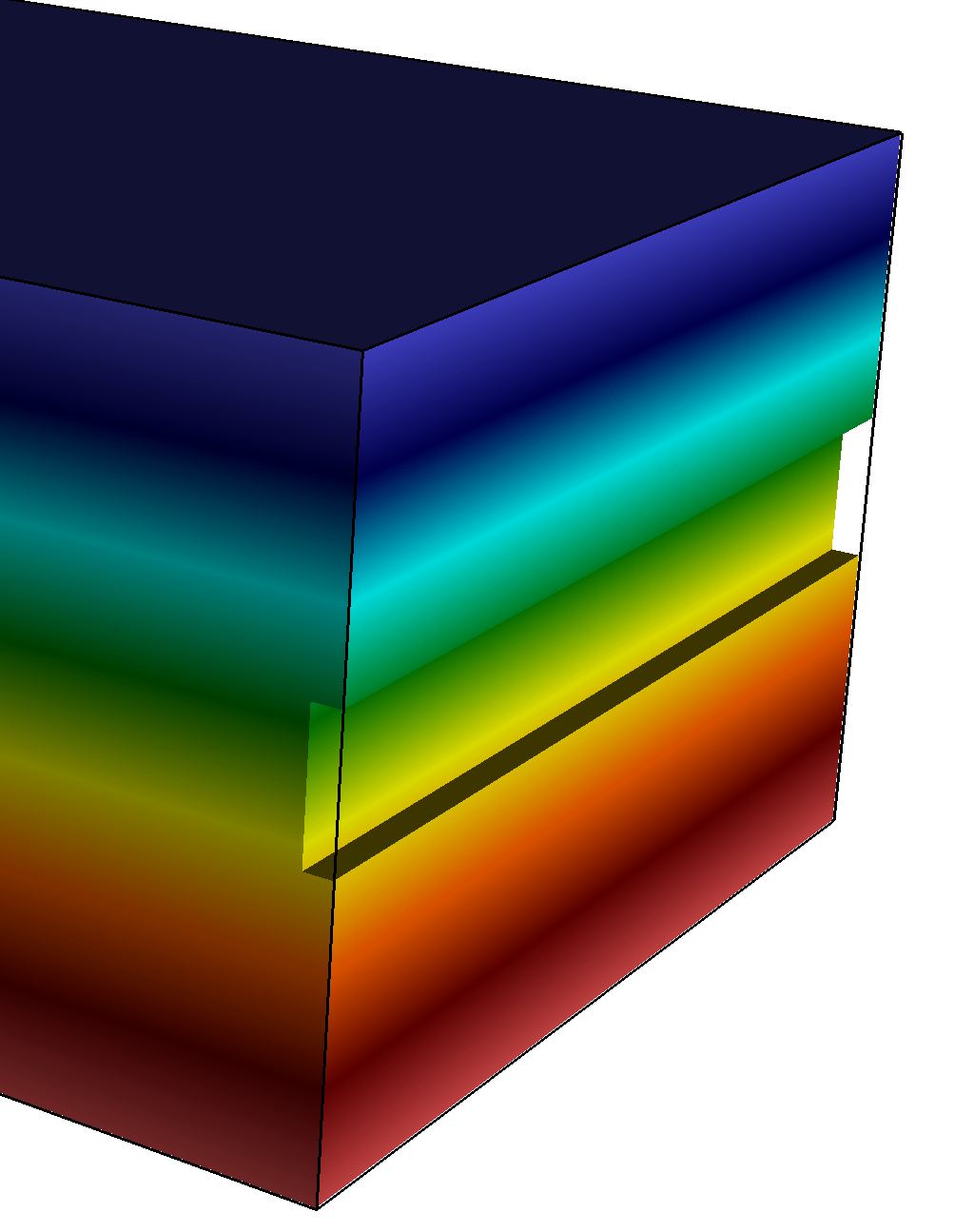}
    \caption{(a) Critical barrier width $\epsilon_c$ as a function of its height $h$ for $E=10^{-4}$ and $Ra=2Ra_c^{\textrm{wall}}\approx6.8\times10^5$. The bottom row shows side visualizations of the temperature field for (b) the case without barrier, (c) $h=0.2$ and $\epsilon=0.02$ and (d) $h=0.2$ and $\epsilon=0.04$. All three cases are indicated using arrows in (a).}
    \label{ec_hb}
\end{figure}

For a given barrier shape defined by $h$ and $\epsilon$, we start the simulation from rest, adding an infinitesimal perturbation to the otherwise linear temperature background. After a short transient, the kinetic energy grows exponentially in time and we measure the associated growth rate. We systematically check that the growing mode does indeed correspond to a wall mode attached to each sidewall. The simulations are repeated for many values of $h$ and $\epsilon$ in order to find the critical curve separating growing from decaying solutions. The results are shown in Fig.~\ref{ec_hb} for both barrier types. In both cases, we find that wall modes can be stabilized for large enough $\epsilon$. The only exception is provided by very tall barriers, $h\rightarrow1$, for which wall modes actually develop along the barrier wall instead of the outer boundary of the domain. Interestingly, for both types of boundary conditions, the critical width $\epsilon_c$ necessary to suppress wall modes tends to a constant when $h\rightarrow0$.

To simplify the system further, we consider an infinitely thin barrier $h\!=\!0$. This is achieved numerically by imposing internal boundary conditions between spectral elements within the fluid domain. In all cases we imposed a no-slip boundary condition on this internal boundary. For the temperature, we required the temperature to be equal to the local equilibrium temperature $T(z)\!=\!1\!-\!z$, or imposed continuity of the temperature and its derivatives as for any other internal spectral element, thereby modeling a thermally conducting barrier. Following the same approach as previously, we found that the critical width of the barrier maintained at fixed temperature is $\epsilon_c\!\approx\!0.047$, a result consistent with the limit observed for finite barriers (empty square in Fig.~\ref{ec_hb}), while for the conducting barrier $\epsilon_c\!\approx\!0.075$ (empty circle), again consistent with the corresponding result for a finite barrier. Thus the barrier height $h$ is a secondary parameter and it is the barrier width $\epsilon$ that is key to suppressing the wall modes.

\begin{figure}
    \centering
    \includegraphics[width=0.4\textwidth]{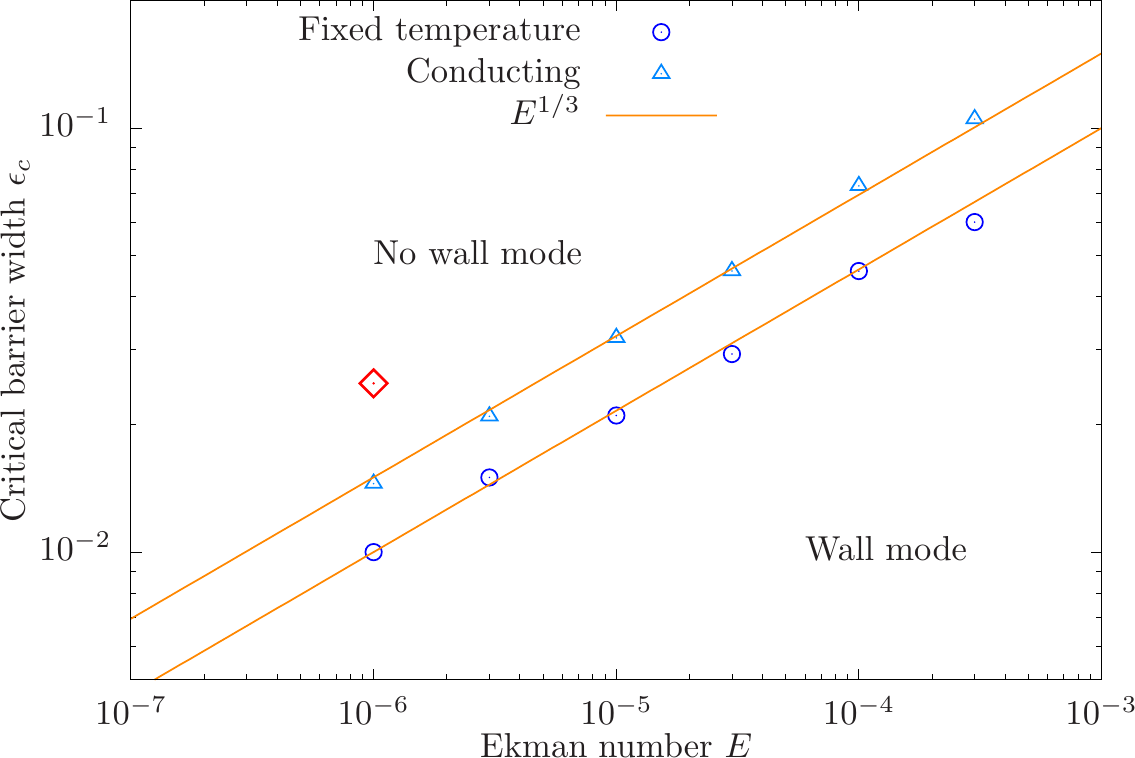}
\caption{Critical barrier width $\epsilon_c$ as a function of $E$ for infinitely thin fixed-temperature or conducting barriers. The Rayleigh number is fixed at twice its onset value, which itself depends on $E$. The red diamond indicates the regime considered for the experimentally relevant simulations shown in Fig.~\ref{fig:nusselt}.}
    \label{ek}
\end{figure}

Let us now discuss the effect of the two main control parameters of the problem, namely $E$ and $Ra$, for an infinitely thin and conducting barrier.
We systematically vary the Ekman number while keeping the Rayleigh number equal to twice its critical value (which itself depends on $E$, thus keeping the supercriticality of the system fixed).
For each Ekman number, we vary the barrier width $\epsilon$ and measure the growth rate in order to approximate its critical value $\epsilon_c$. The results are shown in Fig.~\ref{ek}, where we find that the barrier width required to stabilize wall modes follows a $E^{1/3}$ scaling for both types of temperature boundary conditions, with a prefactor that is slightly larger for the conducting barrier, as already found for $E=10^{-4}$. The $E^{1/3}$ scaling is consistent with the wall mode width at onset \cite{busse1993}.

\begin{figure}
    \centering
    \includegraphics[width=0.4\textwidth]{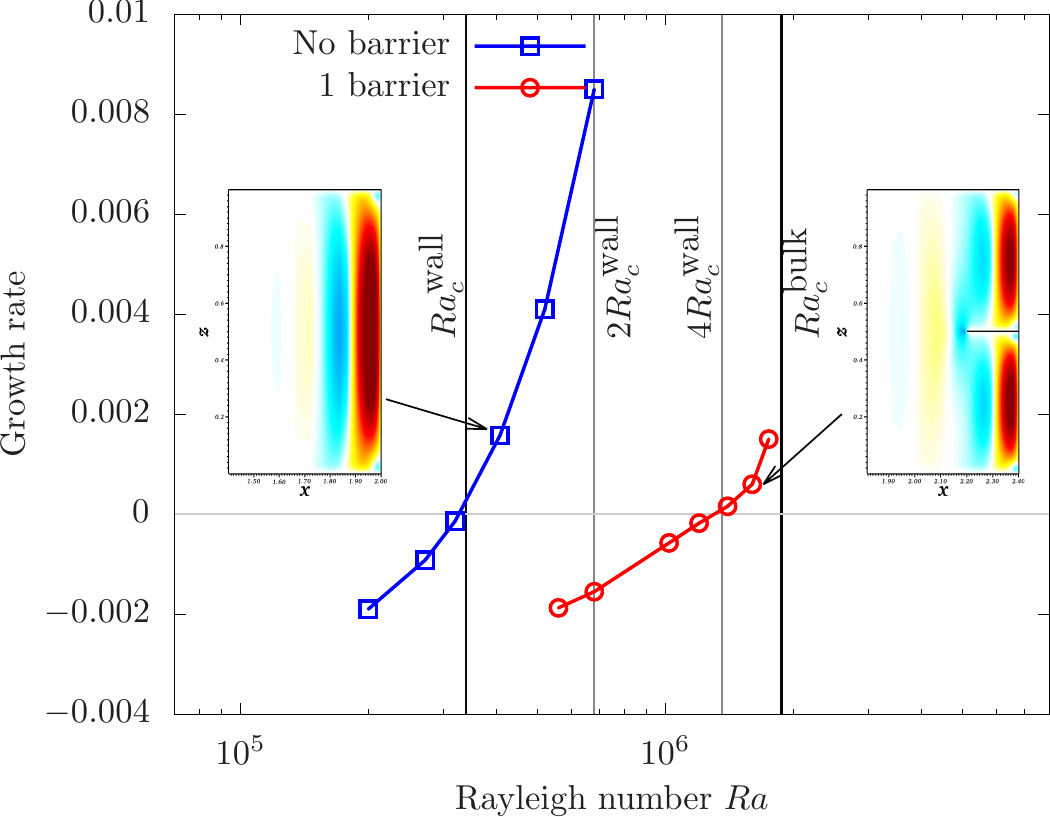}
    \caption{Growth rate as a function of $Ra$ for $E=10^{-4}$ for the reference case with no barrier and the case with one barrier of width $\epsilon=0.2$ located at $z=1/2$. The insets show the vertical component of the velocity in the $y=0$ plane during the exponential growth phase.}
    \label{fig:ra}
\end{figure}

Increasing the Rayleigh number leads to more surprising results.
For a conducting barrier of width $\epsilon=0.2$ at $E=10^{-4}$ the wall modes are suppressed when $Ra=2Ra_c^{\textrm{wall}}$ (Fig.~\ref{fig:ra}).
However, the growth rate increases with $Ra$ so that the wall mode starts to grow at a larger value of $Ra$, which happens at $Ra\approx4Ra_c^{\textrm{wall}}$.
Thus the inclusion of the barrier alters the linear stability of the wall mode by increasing its critical Rayleigh number.
Note that to avoid any potential effect of the barrier width on these results, we have used an unnecessarily wide barrier with $\epsilon=0.2$.
Once the Rayleigh number passes beyond the secondary transition at $Ra\approx4Ra_c$, the growing wall mode cannot be suppressed by a further increase in $\epsilon$.
This is because the wall mode develops on {\it both} sides of the barrier (see the insets in Fig.~\ref{fig:ra}), so that $\epsilon$ becomes irrelevant.
This observation leads to the derivation of a simple model explaining why the wall mode reappears once the Rayleigh number is four times critical. The regions above and below the barrier are controlled by different effective parameters than the bulk: the effective height is reduced by a factor 2 and so is the effective temperature drop. The effective Rayleigh number is thus decreased by a factor 16 while the effective Ekman number is increased by a factor 4. Since the critical Rayleigh number for the wall modes follows a $E^{-1}$ scaling \cite{kuo1993,busse1993}, this means that the effective critical number of the two wall modes developing on either side of the barrier is now increased by a factor 4, as found in Fig.~\ref{fig:ra}.

\begin{figure}
    \centering
    \includegraphics[width=0.48\textwidth]{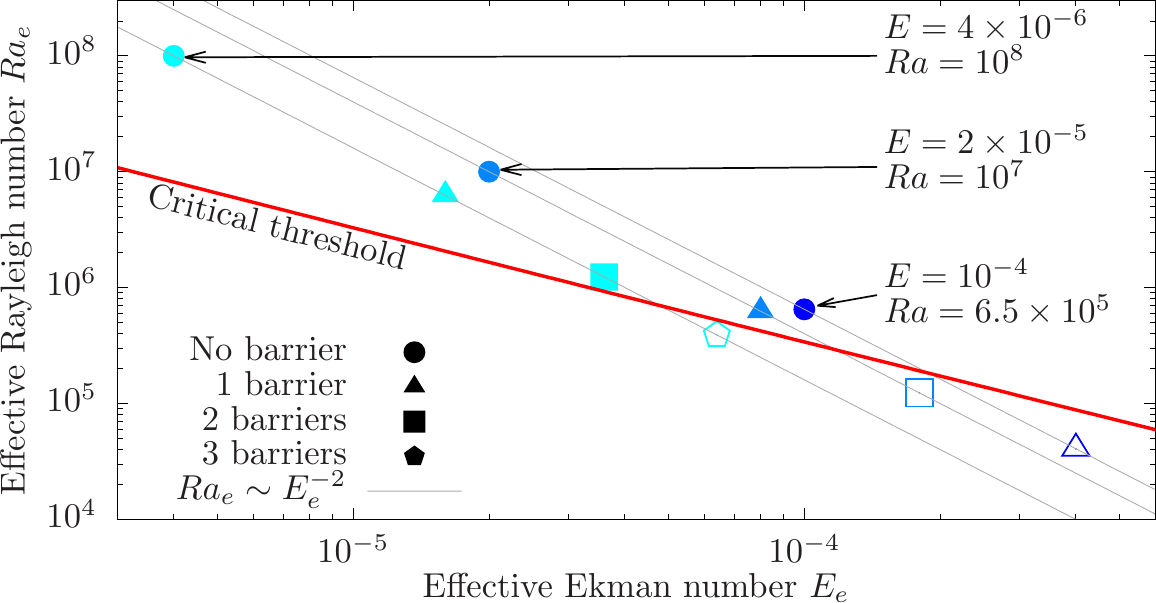}
    \caption{Effective Rayleigh number $Ra_e$ versus the effective Ekman number $E_e$. The symbols indicate the number of barriers starting from $N=0$ where $Ra_e=Ra$ and $E_e=E$. Filled (empty) symbols indicate growing (decaying) wall modes. The thick red line is the critical curve for the onset of wall modes predicted by linear theory \citep{zhang_liao_2009} while the thin grey lines show the $Ra_e\sim E_e^{-2}$ scaling.}
    \label{fig:nbarrier}
\end{figure}

\begin{figure*}
\centering
(a)
\includegraphics[height=0.26\textwidth]{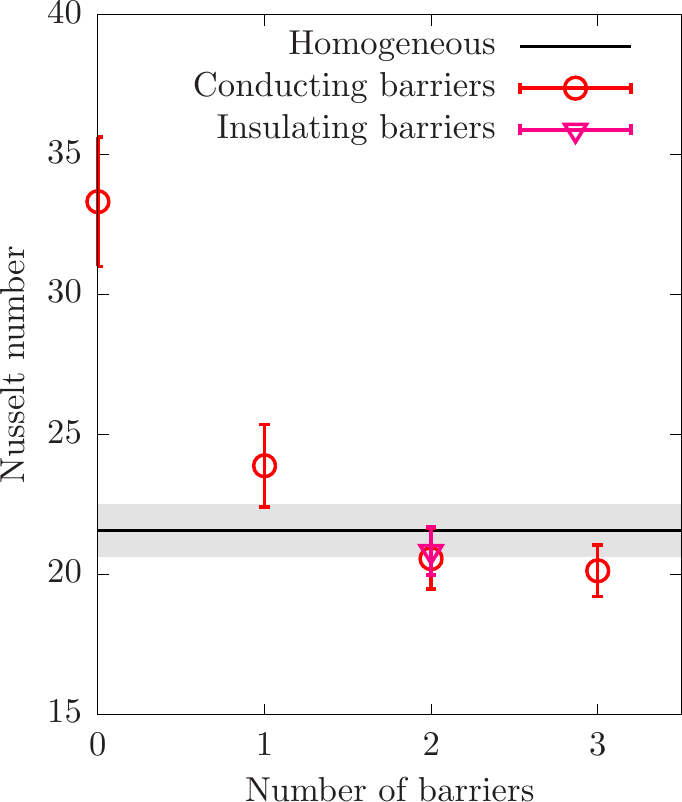}\hfill
(b)
\includegraphics[height=0.26\textwidth]{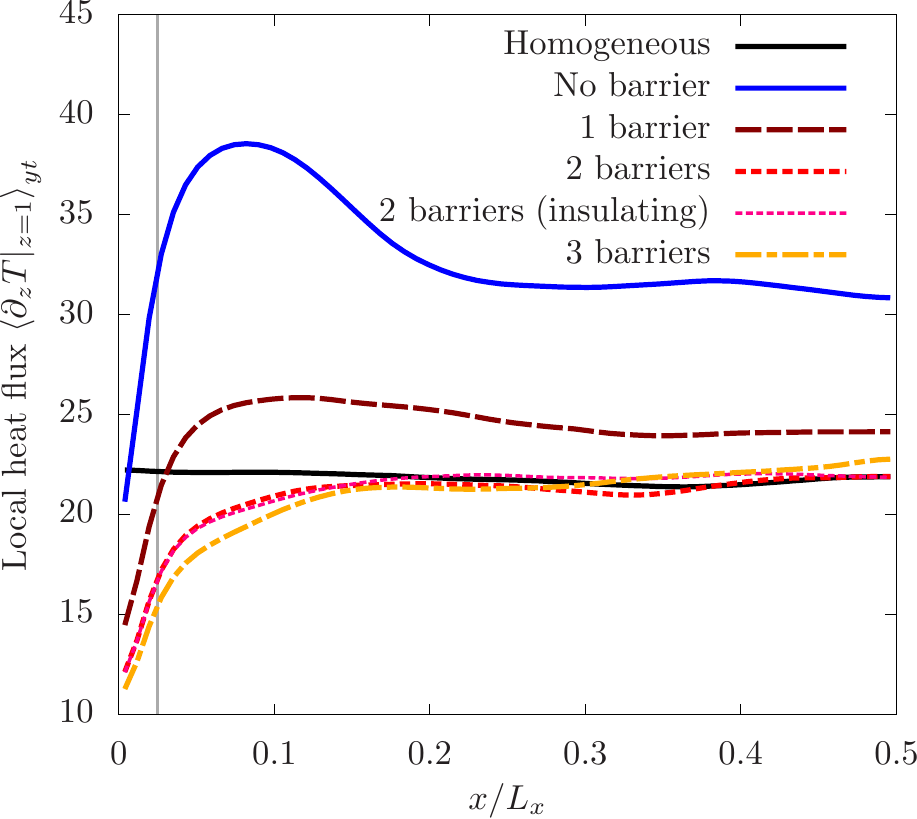}\hfill
(c)
\includegraphics[height=0.28\textwidth]{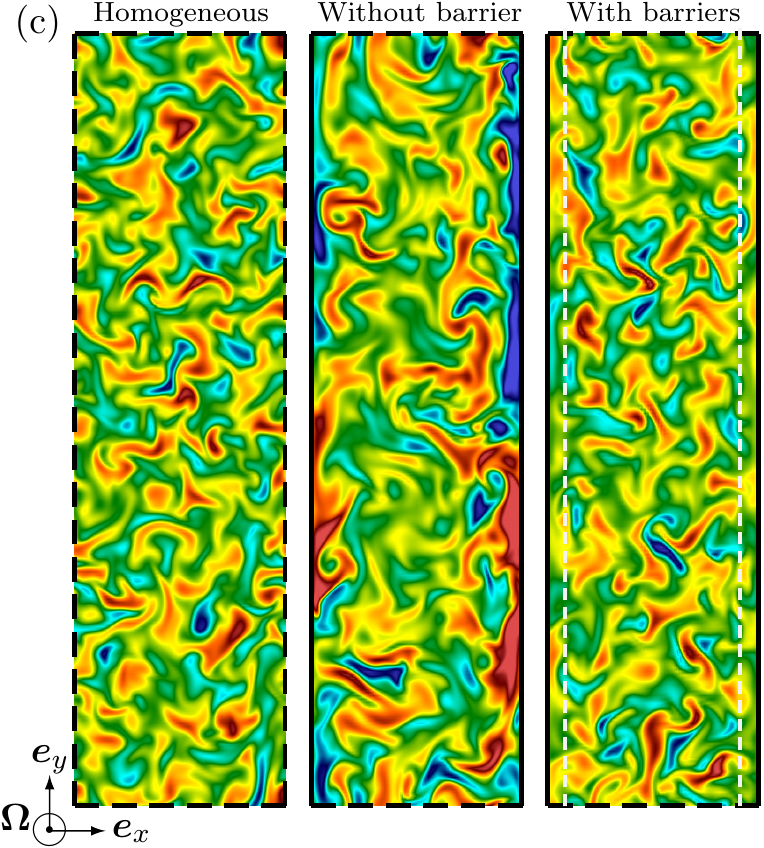}
\caption{(a) Nusselt number as a function of the number $N$ of barriers for $Ra=3\times10^9$ and $E=10^{-6}$. The horizontal line indicates the Nusselt number in the periodic case (the shaded area indicates the 95\% confidence interval). (b) Heat flux at $z=1$ averaged over $y$ and time as a function of the normalized distance from the wall. The vertical line indicates the barrier width. (c) Vertical velocity in the $(x,y)$ plane at $z=1/2$ for the periodic case (left), with vertical walls but no barriers (middle) and with two barriers (right). The same color scale is used in all cases. Black dashed lines indicate periodic boundaries while continuous black lines correspond to rigid walls; the white dashed lines in the interior of the domain show the barrier width. The domain size is $L_x=0.3$ and $L_y=1.1$ \citep{supp}.}
\label{fig:nusselt}
\end{figure*}

One possible solution to the emergence of secondary wall modes as $Ra$ increases is to add more barriers. Assuming that barriers are distributed uniformly along the vertical wall, i.e., that the $N$ barriers are located at $z_i=i/(N+1)$ with $i=1,\dots,N$, we can derive a simple model to predict the minimum number of barriers necessary to stabilize the modes for a given set of $(E,Ra)$. The effective Ekman and Rayleigh numbers for the gap between barriers, with $d$ the distance between two successive barriers and $N$ the number of barriers are
\begin{equation}
E_e\!=\!E d^{-2}\!=\!E(N+1)^2, \ Ra_e\!=\!Ra d^4\!=\!Ra(N+1)^{-4}.
\end{equation}
Thus as more barriers are introduced, $Ra_e \!\sim\! E_e^{-2}$. In order to confirm this simple model, we consider various pairs of control parameters $(,Ra)$. For each case, we ran simulations with a variable number of barriers and fixed $\epsilon\!=\!0.2$. The results are shown in Fig.~\ref{fig:nbarrier} where full (empty) symbols indicate growing (decaying) wall modes. The wall modes disappear when the number of barriers is sufficient to push the effective control parameters $(E_e,Ra_e)$ below the critical curve. Evidently, for $E\!=\!10^{-4}$, $Ra\!=\!6\!\times\!10^5$ one barrier is enough, while for $E\!=\!4\!\times\!10^{-6}$, $Ra\!=\!10^8$ three barriers are necessary in order for each gap to be subcritical.

We now assess the ability of the barriers to damp the wall modes in a fully turbulent environment by reducing the Ekman number to $E=10^{-6}$ while fixing $Ra=3\times10^9\approx4Ra_c^{\textrm{bulk}}$, where $Ra_c^{\textrm{bulk}}$ is the critical Rayleigh number for bulk convection \cite{homsy_hudson_1971}.
This is much closer to the regime relevant to most experimental settings \cite{zhang2021boundary,ecke2022connecting}.
Three types of simulations were performed. In the first periodic boundary conditions were used in both lateral directions to reproduce the geophysical configuration with no walls. The second was a simple channel with sidewalls but no barriers to replicate the experimental configuration. The third included different numbers of barriers. In order to ensure that the barriers were wide enough to affect the wall modes, we chose $\epsilon=0.025$ (red diamond in Fig.~\ref{ek}), which is well above the critical value $\epsilon_c$.
We ran each simulation until a statistically stationary state was reached and computed the Nusselt number by time-averaging the heat flux across the top boundary, typically over $10^4$ rotation periods or more than $500$ free-fall times for our parameters.

The Nusselt number in the confined domain with no barriers is 54\% higher than in the periodic domain [Fig.~\ref{fig:nusselt}(a)], illustrating the dramatic effect of the presence of the wall modes.
Note that the homogeneous flux is not even recovered in the bulk of the domain [see the averaged heat flux as a function of the distance from the wall in Fig.~\ref{fig:nusselt}(b)].
When a single barrier is added, the heat flux decreases drastically, but the Nusselt number for the periodic case is only recovered with two or more barriers.
In fact the Nusselt number falls below the periodic case by approximately 7\%, likely a result of the invasive nature of the barrier, although the homogeneous heat flux is recovered far enough from the wall [Fig.~\ref{fig:nusselt}(b)].
This slight decrease in the total heat flux is smaller than the horizontal surface ratio occupied by the two barriers ($2\epsilon/L_x\approx17$\%) and should decrease as the Ekman number decreases and the wall modes become thinner.
While the wall modes are clearly visible in the middle panel of Fig.~\ref{fig:nusselt}(c), the right panel corresponding to the case with two conducting barriers is virtually indistinguishable from the left panel showing the reference homogeneous case. Further details are provided in the Supplemental Material \cite{supp} where it is shown that the boundary zonal flow is also stabilized while the turbulent fluctuations in the bulk of the homogeneous case are recovered with two or more barriers.
It is of interest that our simple estimate predicts that 8 barriers are required to fully stabilize the wall mode while only two are necessary to recover the effective heat flux in the bulk state, indicating that two barriers reduce the supercriticality of the wall modes sufficiently to render their contribution to the total heat flux negligible even though the wall modes are still presumably present. 


We have shown that the dominant effect of wall modes on convective heat transport in rotating convection experiments can be eliminated by the introduction of one or more thin horizontal fins along the lateral boundary, of order $E^{1/3}$ in width and hence much smaller than the domain radius. This recipe is experimentally realizable and will potentially allow existing and future experiments to realize the state of homogeneous geostrophic turbulence even in the presence of lateral confinement.
Extensive modeling for the proposed experimental parameters along the lines presented here will optimize the required number of barriers and assess the role of the Prandtl number.

\acknowledgments

This work was supported in part by the National Science Foundation under Grant DMS-2009563 (EK).
Centre de Calcul Intensif d'Aix-Marseille is acknowledged for granting access to its high-performance computing resources.
This work was performed using HPC/AI resources from GENCI-IDRIS/TGCC (Grant 2021-A0120407543).

\bibliography{biblio}

\end{document}